\documentclass{aastex}
\usepackage{emulateapj5}

\slugcomment{Version of 11th June, 2007}

\shorttitle{Hercules dSph}
\shortauthors{Coleman et al.}

\begin{document}

\title{The Elongated Structure of the Hercules dSph from Deep LBT Imaging\altaffilmark{1}}

\author{Matthew G.\ Coleman\altaffilmark{2}, Jelte T.\ A.\ De Jong\altaffilmark{2}, Nicolas F.\ Martin\altaffilmark{2}, Hans-Walter Rix\altaffilmark{2}, David J. Sand\altaffilmark{3,4}, Eric F.\ Bell\altaffilmark{2}, Richard W.\ Pogge\altaffilmark{5}, David J.\ Thompson\altaffilmark{6}, H.\ Hippelein\altaffilmark{2}, E.\ Giallongo\altaffilmark{7}, R.\ Ragazzoni\altaffilmark{7}, Andrea DiPaola\altaffilmark{7}, Jacopo Farinato\altaffilmark{8}, Riccardo Smareglia\altaffilmark{9}, Vincenzo Testa\altaffilmark{7}, Jill Bechtold\altaffilmark{3}, John M.\ Hill\altaffilmark{6}, Peter M.\ Garnavich\altaffilmark{10}, Richard F.\ Green\altaffilmark{6}}

\altaffiltext{1}{Based on data acquired using the Large Binocular Telescope (LBT).  The LBT is an international collaboration among institutions in the United States, Italy and Germany. LBT Corporation partners are: The University of Arizona on behalf of the Arizona university system; Istituto Nazionale di Astrofisica, Italy; LBT Beteiligungsgesellschaft, Germany, representing the Max-Planck Society, the Astrophysical Institute Potsdam, and Heidelberg University; The Ohio State University, and The Research Corporation, on behalf of The University of Notre Dame, University of Minnesota and University of Virginia.}
\altaffiltext{2}{Max-Planck-Institut f\"{u}r Astronomie, K\"{o}nigstuhl 17, D-69117 Heidelberg, Germany}
\altaffiltext{3}{Steward Observatory, The University of Arizona, Tucson, AZ 85721}
\altaffiltext{4}{Chandra Fellow}
\altaffiltext{5}{Department of Astronomy, Ohio State University, 140 West 18th Avenue, Columbus, OH 43210-1173}
\altaffiltext{6}{Large Binocular Telescope Observatory, University of Arizona, 933 N. Cherry Ave., Tucson, AZ  85721-0065}
\altaffiltext{7}{INAF, Osservatorio Astronomico di Roma, via Frascati 33, I-00040 Monteporzio, Italy}
\altaffiltext{8}{INAF, Osservatorio Astronomico di Padova, vicolo dell'Osservatorio, 5, 35122 Padova, Italy}
\altaffiltext{9}{INAF, Osservatorio Astronomico di Trieste, via G.B. Tiepolo, 11, 34131 Trieste, Italy}
\altaffiltext{10}{Harvard-Smithsonian Center for Astrophysics, 60 Garden St., Cambridge MA 02138}
\begin{abstract}

We present a deep, wide-field photometric survey of the newly-discovered Hercules dwarf spheroidal galaxy, based on data from the Large Binocular Telescope.  Images in $B$, $V$ and $r$ were obtained with the Large Binocular Camera covering a $23' \times 23'$ field of view to a magnitude of $\sim$25.5 ($5\sigma$).  This permitted the construction of colour-magnitude diagrams that reach approximately 1.5 magnitudes below the Hercules main sequence turnoff.  Three-filter photometry allowed us to preferentially select probable Hercules member stars, and examine the structure of this system at a previously unattained level.  We find that the Hercules dwarf is highly elongated ($3:1$), considerably more so than any other dSph satellite of the Milky Way except the disrupting Sagittarius dwarf.  While we cannot rule out that the unusual structure is intrinsic to Hercules as an equilibrium system, our results suggest tidal disruption as a likely cause of this highly elliptical structure.  Given the relatively large Galactocentric distance of this system ($132 \pm 12$ kpc), signs of tidal disruption would require the Hercules dwarf to be on a highly eccentric orbit around the Milky Way.

\end{abstract}

\keywords{galaxies: individual (Hercules dSph) --- galaxies: kinematics and dynamics}

\section{Introduction}

At the lowest luminosity end of the realm of galaxies, dwarf spheroidal (dSph) galaxies are characterised as low surface brightness systems which are highly dominated by dark matter and have star formation histories ranging from the simple to the complex.  The Milky Way (MW) currently has eighteen known dSph companions, many of these, including the Hercules dwarf spheroidal galaxy \citep{belok07} recently discovered by the Sloan Digital Sky Survey (SDSS; \citealt{sdss}).  Each of these must be experiencing some structural distortion due to the MWs gravitational field, and tidal distortions and disruption have been recognized in some of them: foremost the Sagittarius dSph \citep{ibata94}, but also the Ursa Minor \citep{spick01,palma03} dSph.  However, tidal distortion has not yet been observed in a satellite $>100$ kpc from the Galactic centre, where the tidal forces should be much weaker.  However, it is the pericentric distance of a satellite, not its current distance, that determines how much a satellite is affected by tides.

In this paper, we present and analyse deep images of the recently discovered Hercules dSph, which lies at a distance of $\sim$140 kpc \citep{belok07}.  Belokurov et al.\ inferred that the Hercules dSph displays an extended morphology and may contain multiple stellar populations.  A spectroscopic survey by \citet{simon07} measured a mass-to-light ratio of $332 \pm 221$ in solar units, and found evidence for kinematic substructure.  In this letter, we examine the structure of the Hercules dSph using photometry obtained with the newly commissioned Large Binocular Telescope.  We demonstrate that the structure of Hercules is highly elongated, much more so than all other known distant dSphs, possibly indicating strong tidal distortion.

\section{Deep Photometry from the LBT}
The Large Binocular Telescope (LBT) is located on Mount Graham in Arizona, and consists of two 8.4 metre mirrors on a common mount \citep{hill06}.  Our data were obtained as part of the LBT Science Demonstration Time during which a single mirror of the LBT was fitted with the blue channel of the Large Binocular Camera (LBC; \citealt{rag06,gial07}).  The LBC is a wide-field imager which provides a $23' \times 23'$ field of view, sampled at $0.23$ arcsec/pixel over four chips of $2048 \times 4608$ pixels.  LBC-Blue is optimized for the UV--blue wavelengths, from 320 to 500 nm, and is equipped with the $U$, $B$, $V$, $g$ and $r$ filters.

The Hercules dSph was imaged with the LBT/LBC-Blue setup on March 17 and May 13 2007.  The data consist of five 5 min exposures in Gunn $r$-band, four 5 min exposures in $V$-band and six 5 min exposures in $B$-band (all dithered), taken during photometric conditions with a seeing of approximately $0.8 - 1.1''$.  The data were reduced using standard IRAF routines in the {\em mscred} package: the images on the four chips were trimmed and bias-subtracted, and they were then flat-fielded using the combined twilight flats obtained at the start of the night.  The images contained significant field distortion at the edge, resulting from the `fast' focal ratio ($f$/1.14) of the LBTs primary mirrors.  This was removed before co-adding the science frames, by applying a quadratic radial correction (accuracy $0.2''$).  The Hercules images were then median combined using the {\em mscstack} and {\em mscskysub} routines \citep{valdes02} to produce the final science frames.

Photometry was measured using the PSF-fitting algorithm within the DAOPHOT package.  The photometric dataset was calibrated using stars from SDSS, bypassing any need for an atmospheric extinction correction.  The zero-point uncertainties were $\delta r \sim 0.02$ mag and $\delta B$, $\delta V \sim 0.03$ mag, including the empirical colour transformations described by \citet{jordi06}, without any indication of zeropoint gradients across the LBC field.  We then corrected the photometry for extinction using the dust maps of \cite{schlegel98}, with $\langle E(B - V) \rangle$ of $0.055$ mag.  The photometric uncertainty and completeness as functions of magnitude were assessed using artificial star tests.  We placed 1600 artificial stars in the image and attempted to recover them with DAOPHOT, where the photometric uncertainty was then determined as the dispersion of the returned magnitudes about the mean (that is, not the input).  This was repeated for artificial stars at every 0.25 magnitudes in the $B$, $V$ and $r$ frames.  

The resulting CMDs for selected regions of the image are shown in Fig.\ \ref{herccmd}.  For our analysis of the Hercules dSph, we adopted the specially fit `colours', $c_1$ and $c_2$, which represent a combination of our three filter photometry in $B$, $V$ and $r$.  These principal colours are designed to enhance the Hercules-to-field contrast in the CMD (see \S \ref{structure}), and are similar to those used by \citet{oden01}.  By fitting a ridge-line locus in this plane using the stars at the centre of Hercules, we found the principal colours to be: $c_1 = 0.944(B - V) + 0.330(V - r)$, $c_2 = -0.330(B - V) + 0.944(V - r)$.  A diagram in the $(c_2,V)$ plane is essentially the CMD seen edge-on, and the dispersion of Hercules member stars around $c_2=0$ is defined by the photometric errors.  For the remainder of this paper, our CMDs will be defined in the $(c_1, V)$ plane.  The blue errorbars in Fig.\ \ref{herccmd} represent the combined uncertainty of the colour $c_1$, and the dashed lines represent the magnitude at which the photometric completeness has fallen to $50\%$: $B_{50}=25.60 \pm 0.05$ mag, $V_{50}=25.72 \pm 0.05$ mag, $r_{50}=25.56 \pm 0.04$ mag.

\section{The Structure of Hercules} \label{structure}

The CMD of all stellar sources in the Hercules core region is shown in Fig.\ \ref{herccmd}, the first data for this galaxy to reach well below the main sequence turnoff.  From the $B$ and $V$ photometry we derived the best-fitting distance, age and metallicity for a single stellar population (Table \ref{hercpars}) using the CMD-fitting techniques described by \citet{dejong07} and originally developed by \citet{dolphin02}.  We find Hercules to be dominated by old and metal-poor stars.  The dashed line in Fig.\ \ref{herccmd} traces the isochrone for a stellar population with an [Fe/H] abundance of $-2.26$ and an age of 13 Gyr \citep{girardi02,girardi04}.  A detailed analysis of the star formation history and metallicity evolution of Hercules will be presented in a later publication.

Starting from the CMD of all detected sources in the entire field, we tried to isolate probable Hercules dSph stars by excising ``field sources'', such as foreground stars and background galaxies, using a CMD selection technique originally introduced by \citet{grillmair95}; see also \citealt{oden01}): a comparison was made between the $(c_1,V)$ CMD of the central region of Hercules and that of the field region (a distant portion of the $23' \times 23'$ field of view) to produce a `signal-to-noise' map of the Hercules stellar population across the CMD.  The CMDs for the core of Hercules and the field are shown in Fig.\ \ref{herccmd}, where the associated spatial regions were chosen iteratively and are outlined in Fig.\ \ref{hercxy}.  To make a map of probable Hercules member stars (Fig.\ \ref{hercxy}), we then chose a limiting signal-to-noise in the CMD plane ($s = 3.0$, the red contour in Fig.\ \ref{herccmd}) which minimised the field population while maintaining a high number of Hercules stars.  This corresponds to the main sequence turnoff and upper main sequence region of the Hercules CMD; only stars above the $50\%$ completeness limit shown in Fig.\ \ref{herccmd} were selected.

At such faint magnitudes, background galaxies form the majority of contaminating sources, consequently optimal star-galaxy separation is important.  The FWHM of many background galaxies are expected to be significantly larger than that of a star; hence, the DAOPHOT `sharpness' parameter provided a secondary level of filtering.  We used the sharpness values from the $B$-band image, for which the PSF shape was effectively constant across the field and the seeing was best ($0.8''$).  We constructed a sharpness map across the LBT $B$-band image using the bright stars ($14 \le B \le 22$) and removed all sources with a non-stellar sharpness.  Together, the CMD selection and star-galaxy separation techniques removed $85\%$ of all stars, most of them from the field population.

The spatial distribution of the `cleaned' Hercules dataset is shown in Fig.\ \ref{hercxy} (upper panel).  A relatively bright star ($g = 15.5$; SDSS) near the centre of the Hercules dSph is saturated in our data.  The data in this region were excluded from our analysis.  We converted this map into a stellar surface density contour diagram (Fig.\ \ref{hercxy}, lower panel) by convolving each star with a Gaussian of radius $0.6'$ (independent of stellar magnitude), where the contours trace stellar densities of $1.5\sigma, 3\sigma,\dots,10.5\sigma$ above the field density, where $\sigma$ is the variance of the mean background level within a circle of radius $0.6'$.

Both panels in Fig.\ \ref{hercxy} show a strikingly elongated stellar distribution of the Hercules dSph.  We quantified the internal structure of Hercules by fitting a series of radially increasing elliptical contours to the stellar density map displayed in Fig.\ \ref{hercxy} using the IRAF routine {\it ellipse}.  The best-fitting position angle, ellipticity and central coordinates were measured at semi-major radii of $0.25', 0.75',\dots, 9.75'$ for 200 random data subsets ($50\%$ of all stars), with associated uncertainties calculated using bootstrap-sampling.  The results in Fig.\ \ref{radial} indicate that the ellipticity of Hercules is relatively mild towards the centre\footnote{Our measurement for the ellipticity of the inner regions is likely to be a lower limit due to the `blurring' effect of the Gaussian convolution function described above.  Also, the ellipticity is measured in the plane of the sky, and Hercules may be further elongated along the line of sight.} ($e \sim 0.3$), however this increases sharply to an ellipticity of $e \sim 0.65$ in the outer regions.  That is, the major-to-minor axis ratio of Hercules is approximately $3:1$, flatter than any other known Milky Way dSph except the tidally disrupted nearby Sagittarius dSph.

We also constructed an azimuthally averaged stellar density radial profile of the dSph in a series of concentric annuli (separated by $0.5'$ with a fixed ellipticity of $e = 0.65$ and position angle of $-73^{\circ}$).  The background level was calculated as the density of stars in the upper ($\Delta \delta \ge 8'$) and lower ($\Delta \delta \le -5'$) regions of the LBT field.  This was subtracted from the profile datapoints, and the result is shown in Fig.\ \ref{radial}.  We derived a best-fitting \citet{king62} profile for this system using bootstrap-resampling, yielding a half-light (major axis) radius of $4.37 \pm 0.29$ arcmin ($168 \pm 11$ pc), a core radius of $4.74 \pm 0.57$ arcmin ($182 \pm 22$ pc), and a tidal radius of $25.9 \pm 11.1$ arcmin ($994 \pm 426$ pc).  Note that the formal tidal radius of the Hercules dSph is beyond the spatial range of our dataset, and should be treated with caution.

\section{Discussion}
In their discovery paper, \citet{belok07} examined the Hercules structure based on photometry complete to $i \sim 22.5$ (the depth of the sub-giant branch) and found that the system has an extended elongated morphology.  In our analysis, we have probed $\sim$3 magnitudes deeper, allowing a structural map of much greater signal-to-noise.  Our data clearly confirm that this object is highly elongated ($3:1$) and shows some tentative indication of clumpy sub-structure\footnote{Note that the visual impression of sub-structure in Fig.\ \ref{hercxy} is affected by the presence of the bright star which is represented by the blue hashed region.} (see Fig.\ \ref{hercxy}).

It is possible that the elongated structure of Hercules is intrinsic to this system.  If so, it would be the most flattened of all the Milky Way dSphs (except the Sagittarius dSph).  All other known stellar systems with $3:1$ flattening or more are predominately supported by rotation; yet, no stellar system with a characteristic velocity as low as Hercules ($\sigma \sim 5$ km s$^{-1}$; \citealt{simon07}) is known to be rapidly rotating.  If in equilibrium, Hercules would have unique structural or kinematic properties.

Structurally, the Hercules dSph most closely resembles the Ursa Minor dSph ($e \sim 0.56$), a Milky Way satellite that is presumably disrupting \citep{spick01,palma03}.  The only other highly elliptical Milky Way dSph is the Sagittarius system, a relatively bright dSph that is being disrupted by Galactic tidal forces.  Tidal forces could therefore be a plausible explanation for the enlongated Hercules structure.  However, the large distance of Hercules is problematic for this interpretation.  Tidal forces decrease as $R^{-3}$, and the presence of tidal distortion would imply that Hercules is on a highly elliptical orbit around the Milky Way, to allow the system to experience strong tidal forces during its pericentric passage.

A simple condition for tidal disruption of Hercules at its orbital pericentre, $R_{\mbox{\scriptsize peri}}$, is $\sigma_{\mbox{\scriptsize Herc}}/r_{\mbox{\scriptsize Herc}} \approx \sigma_{\mbox{\scriptsize MW}}/R_{\mbox{\scriptsize peri}}$, where $\sigma_{\mbox{\scriptsize Herc}}$ and $\sigma_{\mbox{\scriptsize MW}}$ are the velocity dispersions of the Hercules dSph and the Milky Way respectively, $r_{\mbox{\scriptsize Herc}}$ is the limiting radius of Hercules.  For $r_{\mbox{\scriptsize Herc}} \sim 250$ pc (that is, the geometric mean of the major and minor axes), $\sigma_{\mbox{\scriptsize Herc}} \sim 5.1$ km s$^{-1}$ \citep{simon07}, and $\sigma_{\mbox{\scriptsize MW}} \sim 150$ km s$^{-1}$, we find that only $R_{\mbox{\scriptsize peri}} \sim 8$ kpc would lead to disruption.  Given that Hercules is now located at a distance of 132 kpc (and may not yet be at apogalacticon) this would imply an orbit for this system of extreme ellipticity, $e > 0.90$.  A detailed analysis of the feasibility and plausibility of this scenario is warranted, but beyond the scope of this letter.  At any rate, regardless of whether Hercules' highly flattened structure is due to tides, which appears on balance most plausible, or intrinsic to an equilibrium system, its properties are exceptional.

\acknowledgments
The authors thank the LBT Science Demonstration Time (SDT) team for assembling and executing the SDT program.  We also thank the LBC team and the LBTO staff for their kind assistance.  M.C.\ acknowledges the assistance of Roland Gredel when accessing LBT data, and that of Tom Herbst in the preparation of this letter.  The authors thank Dennis Zaritsky and Matthias Steinmetz for their helpful comments on the manuscript.

\plotone{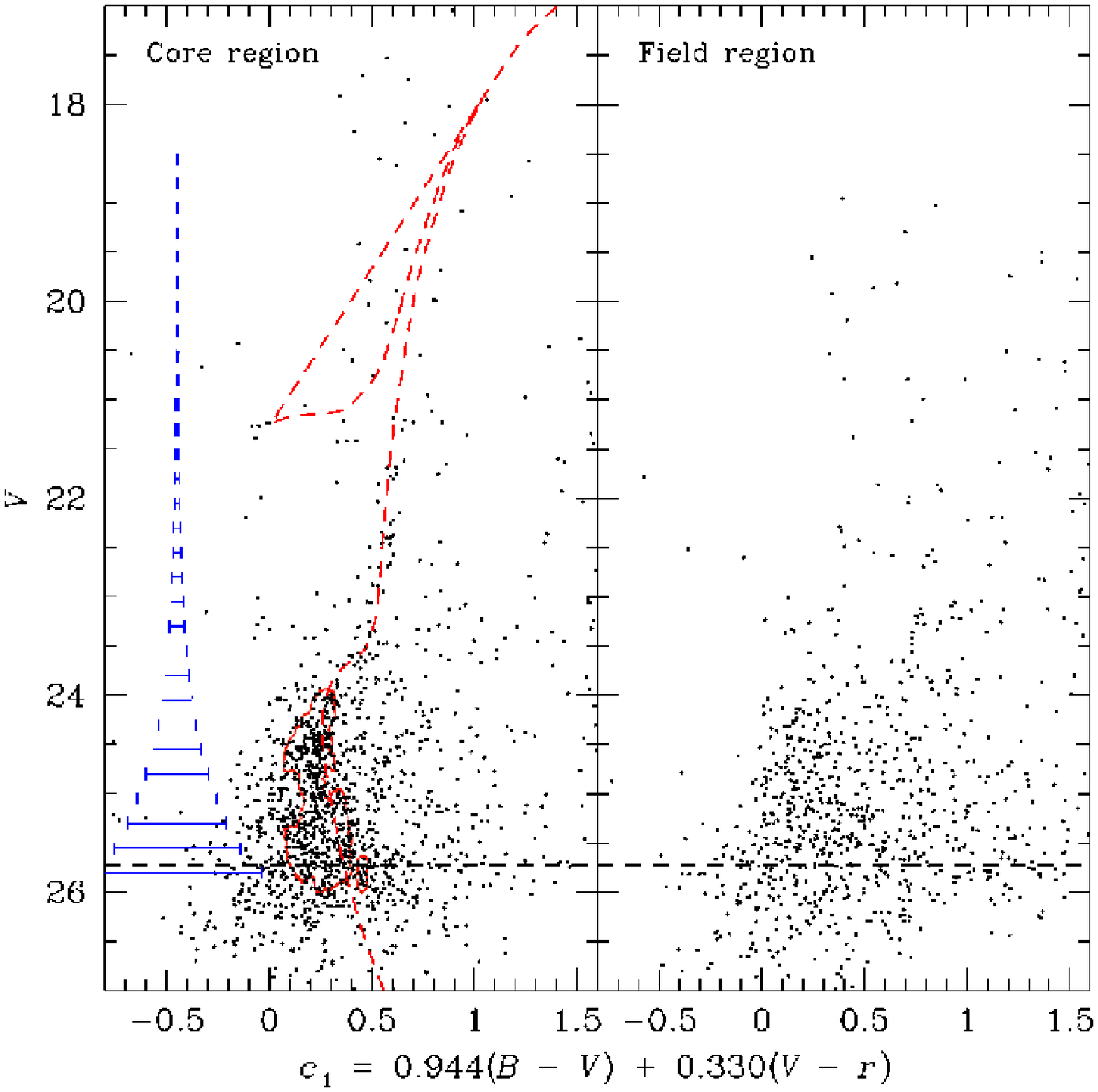}
\figcaption[Hercules CMD]{Colour-magnitude diagrams for the central region of Hercules (left panel) and a field region of equivalent area on the sky (right panel).  The photometry has been corrected for extinction and all sources outside the sharpness limit have been removed.  Our principal colour, $c_1$, has been designed to enhance the Hercules-to-field contrast in the colour-magnitude plane.  The red dashed line is the isochrone of a metal-poor stellar population at 13 Gyr \citep{girardi02,girardi04}.  The blue errorbars in the left panel represent the uncertainty in the colour $c_1$ returned by the artificial star tests, and the dashed lines trace the $V$ magnitude at which the photometric completeness has fallen to $50\%$. The red contour line traces the CMD-selection region for the Hercules main sequence population.  Note the increased background galaxy contamination in the right panel at magnitudes fainter than $V=24$. \label{herccmd}}

\plotone{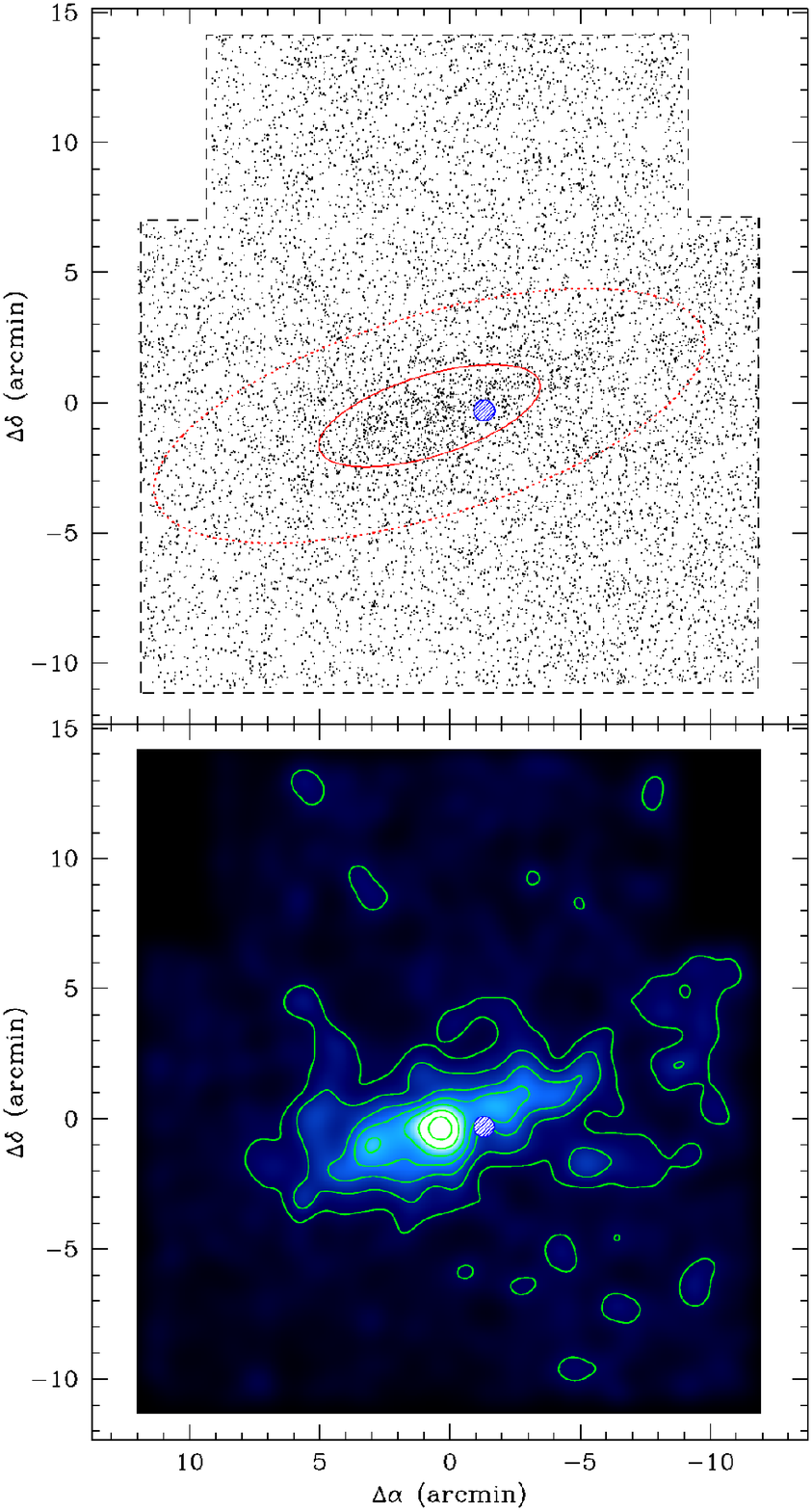}
\figcaption[Spatial distribution of the Hercules CMD-selected sources]{{\em Upper panel:} Spatial distribution of the Hercules CMD-selected objects, where the dashed line marks the limit of our LBT data.  The blue shaded region represents the saturated star aligned towards the centre of the dSph.  The solid red ellipse marks the core radius from our best-fitting King model (see Fig.\ \ref{radial}) and the dashed ellipse outlines the inner border of the field region.  Both the core and field populations were used to derive the CMD-selection limit shown in Fig.\ \ref{herccmd}.  These ellipses have an ellipticity of 0.65 and semi-major axis radii of $4.42'$ and $11'$ respectively.  {\em Lower panel:} Contour diagram of the CMD-selected sources.  Each star has been convolved with a Gaussian of width $0.6'$ arcmin.  The contours correspond to stellar densities of $1.5\sigma, 3\sigma,\dots,10.5\sigma$ above the background, where $\sigma$ is the uncertainty in the background stellar density from Poisson statistics.  \label{hercxy}}

\plotone{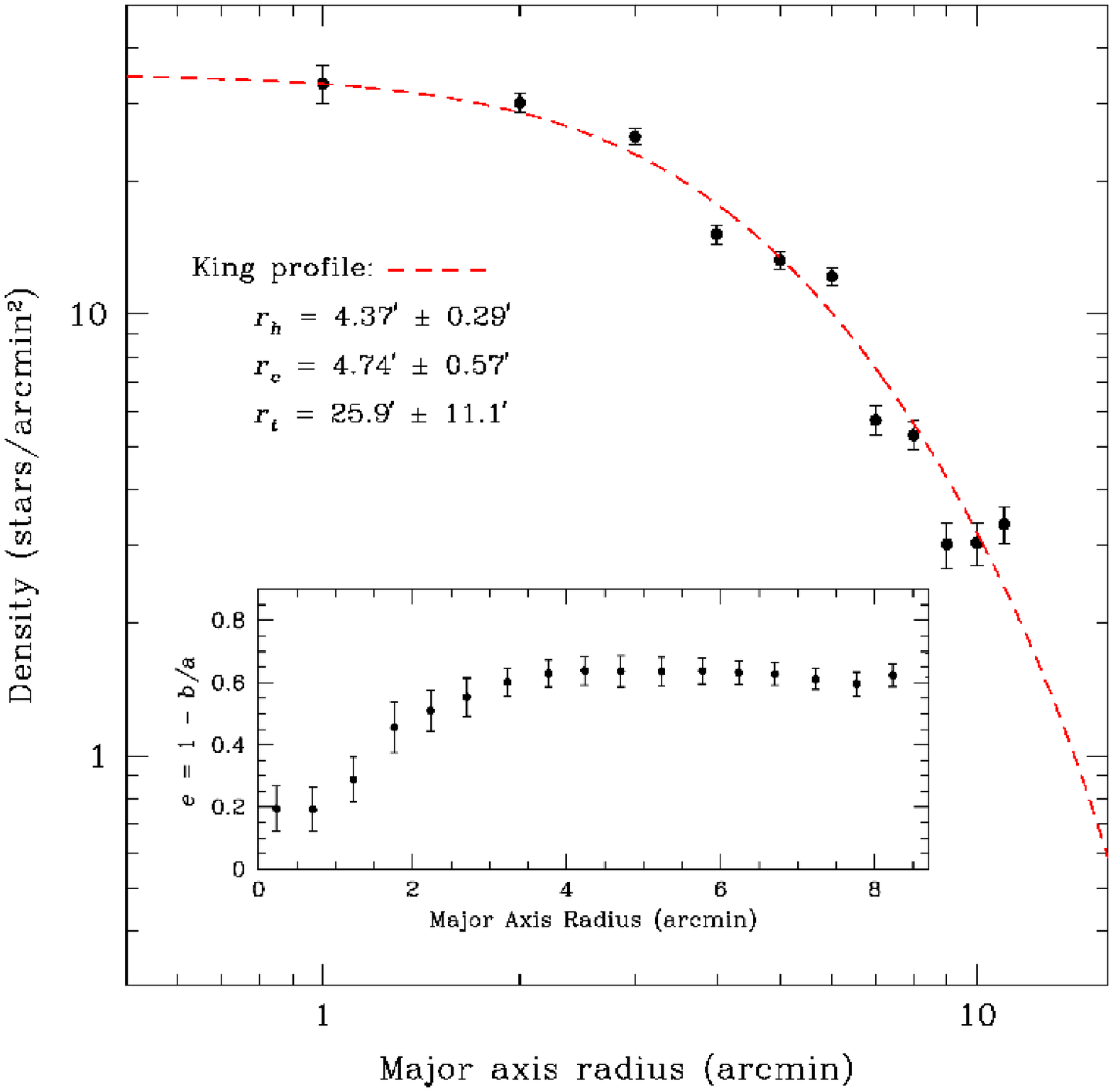}
\figcaption[Hercules radial profile]{Radial profile of Hercules, determined from the CMD-selected dataset.  The stellar density was evaluated within ellipses at every $1'$ major axis radius using an ellipticity of 0.65 and a position angle of $-73^{\circ}$.  The dashed line represents the best-fitting King profile, and the parameters with their associated bootstrap uncertainties are listed.  The background level of $15.43 \pm 0.28$ stars/arcmin$^2$ (Poisson statistical uncertainty) has been subtracted from all datapoints.  {\em Inset panel:} Ellipticity as a function of radius.  Uncertainties were determined using bootstrap resampling.  The central coordinates of this system demonstrate a mild dependence on radius, with a variation of $\sim$$0.3'$ in RA and $\sim$$0.2'$ in Dec over the radial range. \label{radial}}


\begin{table}
\begin{center}
\caption{Properties of the Hercules dSph}
\label{hercpars}
\vspace{0.2cm}
\begin{tabular}{lc}
\tableline
\tableline
Parameter & Value \\
\tableline
R.A. (J2000) & 16:31:02.0 \\
Decl. (J2000) & 12:47:29.6 \\
$E(B-V)$ (mag) & $0.055 \pm 0.005$\tablenotemark{a} \\
$(m - M)_0$ (mag) & $20.6 \pm 0.2$ \\
Distance (kpc) & $132\pm 12$ \\
$[\mbox{Fe}/\mbox{H}]$ & $-2.1 \pm 0.2$ \\
Age (Gyr) & $13 \pm 3$ \\
 & \\
King $r_h$ & $4.37' \pm 0.29'$ ($168 \pm 11$ pc) \\
King $r_c$ & $4.74' \pm 0.57'$ ($182 \pm 22$ pc) \\
King $r_t$ & $25.9' \pm 11.1'$ ($994 \pm 426$ pc) \\
$c=\log{(r_t/r_c)}$ & $0.74 \pm 0.25$ \\
\tableline
\end{tabular}
\end{center}
\tablenotetext{a}{From \citet{schlegel98}.  The uncertainty quoted here represents the variation in reddening over the LBT field.}
\end{table}



\begin{thebibliography}{}

\bibitem[Adelman-McCarthy et al.(2007)]{sdss} Adelman-McCarthy, J.~K., et al., \apjs, in press
\bibitem[Belokurov et al.(2007)]{belok07} Belokurov, V., et al.\ 2007, \apj, 654, 897
\bibitem[de Jong et al.(2007)]{dejong07} de Jong, J.~T.~A., Butler, D.~J., Rix, H.-W., Dolphin, A.~E., \& Martinez-Delgado, D.\ 2007, ArXiv Astrophysics e-prints, arXiv:astro-ph/0701140 
\bibitem[Dolphin(2002)]{dolphin02} Dolphin, A.~E.\ 2002, \mnras, 332, 91 
\bibitem[Giallongo et al.(2007)]{gial07} Giallongo, E., Ragazzoni, R., Grazian, A. et al. 2007 in preparation
\bibitem[Girardi et al.(2002)]{girardi02} Girardi, L., Bertelli, G., Bressan, A., Chiosi, C., Groenewegen, M.~A.~T., Marigo, P., Salasnich, B., \& Weiss, A.\ 2002, \aap, 391, 195 
\bibitem[Girardi et al.(2004)]{girardi04} Girardi, L., Grebel, E.~K., Odenkirchen, M., \& Chiosi, C.\ 2004, \aap, 422, 205
\bibitem[Grillmair et al.(1995)]{grillmair95} Grillmair, C.~J., Freeman, K.~C., Irwin, M., \& Quinn, P.~J.\ 1995, \aj, 109, 2553
\bibitem[Hill et al.(2006)]{hill06} Hill, J.~M., Green, R.~F., \& Slagle, J.~H.\ 2006, \procspie, 6267, 62670Y
\bibitem[Ibata, Gilmore, \& Irwin(1994)]{ibata94} Ibata, R.~A., Gilmore, G., \& Irwin, M.~J.\ 1994, Nature, 370, 194 
\bibitem[Jordi et al.(2006)]{jordi06} Jordi, K., Grebel, E.~K., \& Ammon, K.\ 2006, \aap, 460, 339 
\bibitem[King(1962)]{king62} King, I.\ 1962, \aj, 67, 471
\bibitem[Mart{\'{\i}}nez-Delgado et al.(2001)]{spick01} Mart{\'{\i}}nez-Delgado, D., Alonso-Garc{\'{\i}}a, J., Aparicio, A., \& G{\'o}mez-Flechoso, M.~A.\ 2001, \apjl, 549, L63 
\bibitem[Odenkirchen et al.(2001)]{oden01} Odenkirchen, M., et al.\ 2001, \aj, 122, 2538 
\bibitem[Palma et al.(2003)]{palma03} Palma, C., Majewski, S.~R., Siegel, M.~H., Patterson, R.~J., Ostheimer, J.~C., \& Link, R.\ 2003, \aj, 125, 1352
\bibitem[Ragazzoni et al.(2006)]{rag06} Ragazzoni, R., et al.\ 2006, \procspie, 6267, 626710
\bibitem[Schlegel et al.(1998)]{schlegel98} Schlegel, D.~J., Finkbeiner, D.~P., \& Davis, M.\ 1998, \apj, 500, 525
\bibitem[Simon \& Geha(2007)]{simon07} Simon, S.~D., Geha, M.\ 2007, \apj, submitted, astro-ph/0706.0516
\bibitem[Valdes(2002)]{valdes02} Valdes, F.~G.\ 2002, Automated Data Analysis in Astronomy, 309 

\end{thebibliography}
\end{document}